\documentclass[conference]{IEEEtran}
\usepackage{cite}
\usepackage{amsmath,amssymb,amsfonts}
\usepackage{algorithmic}
\usepackage{graphicx}
\usepackage{textcomp}
\usepackage[dvipsnames]{xcolor}
\usepackage{tikz}
\usetikzlibrary{quantikz2}
\usepackage[short]{optidef}
\usepackage{orcidlink}
\usepackage{comment}
\usepackage{url}

\DeclareMathOperator*{\argmin}{arg\,min}

\makeatletter
\newcommand{\linebreakand}{%
  \end{@IEEEauthorhalign}
  \hfill\mbox{}\par
  \mbox{}\hfill\begin{@IEEEauthorhalign}
}
\makeatother

\def\BibTeX{{\rm B\kern-.05em{\sc i\kern-.025em b}\kern-.08em
    T\kern-.1667em\lower.7ex\hbox{E}\kern-.125emX}}
\begin{document}

\title{
Portfolio construction using a sampling-based variational quantum scheme
}

\author{
\IEEEauthorblockN{Gabriele Agliardi \orcidlink{0000-0002-1692-9047}}
\IEEEauthorblockA{
\textit{IBM Quantum}\\
\textit{IBM Research -- Italy}\\
}
\and
\IEEEauthorblockN{Dimitris Alevras}
\IEEEauthorblockA{
\textit{IBM Quantum}\\
\textit{IBM Research -- US}
}
\and
\IEEEauthorblockN{Vaibhaw Kumar}
\IEEEauthorblockA{
\textit{IBM Quantum}\\
\textit{IBM Research -- US}
}
\and
\IEEEauthorblockN{Roberto Lo Nardo}
\IEEEauthorblockA{
\textit{IBM Quantum}\\
\textit{IBM Research -- UK}
}
\linebreakand
\IEEEauthorblockN{Gabriele Compostella}
\IEEEauthorblockA{
\textit{IBM Quantum}\\
\textit{IBM Research -- Germany}
}
\and
\IEEEauthorblockN{Sumit Kumar}
\IEEEauthorblockA{
\textit{IBM Quantum}\\
\textit{IBM Research -- US}
}
\and
\IEEEauthorblockN{Manuel Proissl \orcidlink{0009-0006-1083-1133}}
\IEEEauthorblockA{
\textit{IBM Quantum}\\
\textit{IBM Research -- CH}
}
\and
\IEEEauthorblockN{Bimal Mehta}
\IEEEauthorblockA{
\textit{Centre for Analytics \& Insights}\\
\textit{Vanguard}
}
}

\newcommand{\vk}[1]{\textcolor{Green}{[VK] #1}}
\newcommand{\da}[1]{\textcolor{Red}{[DA] #1}}

\maketitle
\begin{abstract}
The efficient and effective construction of portfolios that adhere to real-world constraints is a challenging optimization task in finance. We investigate a concrete representation of the problem with a focus on design proposals of an Exchange Traded Fund. We evaluate the sampling-based CVaR Variational Quantum Algorithm (VQA), combined with a local-search post-processing, for solving problem instances that beyond a certain size become classically hard. We also propose a problem formulation that is suited for sampling-based VQA. Our utility-scale experiments on IBM Heron processors involve 109 qubits and up to 4200 gates, achieving a relative solution error of 0.49\%. Results indicate that a combined quantum-classical workflow achieves better accuracy compared to purely classical local search, and that hard-to-simulate quantum circuits may lead to better convergence than simpler circuits. Our work paves the path to further explore portfolio construction with quantum computers.

\end{abstract}


\section{Introduction}

Portfolio optimization is a systematic mathematical method for selecting investment choices among a set of financial instruments or assets. Portfolios are selections from a possible set of assets and are conventionally defined by their weights, while portfolio optimization methods can also utilize holdings. The common textbook approach, also referred to as modern portfolio theory proposed by Markowitz~\cite{markowitz_portfolio_1952}, entails classifying the investment choices according to risk (standard deviation) and return, thereafter, selecting the portfolio of investments that attains a preferred risk-return tradeoff. Portfolios that meet these characteristics are classified as efficient portfolios, and the graphical representation of their risks and returns delineates a curve known as the efficient frontier. The Markowitz model, despite its conciseness and elegance, contains strong simplifications~\cite{kolm_portfolio_2014} compared to the actual conditions that exist in a financial market. More sophisticated approaches are typically too computationally intensive. Additionally, including stochasticity or working with a multi-period problem is particularly more challenging~\cite{koch_optimization_2025}. Moreover, the Markowitz model is very sensitive to errors in risk-return estimators forming the input data and thus the efficient frontier can shift considerably even with small changes in the inputs.

Conceptually, portfolio optimization theory has vast applications in various flavors for portfolio construction and diversification tasks. A prominent domain is the creation of Exchange Traded Funds (ETFs), where the respective strategy may entail the construction of a portfolio that tracks an existing index of investments or basket while optimally targeting an allocation of investment instruments that maximize returns and minimize risks, while maintaining several investment and compliance related constraints. However, these common objectives are often extended with others that also optimize with respect to funding cost, shot selling, portfolio rebalancing cost etc. and yield an exceedingly more complex optimization problem. Also, as the number of candidates, assets or financial instruments, in the considered investment universe increases, a substantially larger search space leads to non-linear growth of the respective optimization problem complexity.

Portfolio creation of ETFs involves the ideal composition of primary market assets to mimic or anticipate secondary market behavior~\cite{cornuejols2018optimization, dembo1999practice}. The portfolio is often generated by executing \textit{what-if} scenario simulations with user-defined market sensitivity. Optimization-based simulations must be rapid so investment professionals can iterate before making more informed choices. Realistic problems may involve 1,000 to 10,000 candidate instruments. At this problem size, professional classical solvers struggle to provide good solutions at the order of 5–10 minutes~\footnote{Empirical estimate based on internal work}. The vast range of challenges that classical algorithms and computing systems have in efficiently and effectively solving such optimization problems motivate the exploration of quantum technology that aim to improve these real-world workflows.

In this paper, we investigate the feasibility of a quantum-classical workflow in the context of solving a specific bond ETF portfolio construction problem using the Conditional Value at Risk-based Variational Quantum Algorithm (CVaR-VQA)~\cite{barkoutsos_improving_2020}. Recently, CVaR-VQA~\cite{barkoutsos_improving_2020, kumar2025towards, sharma2025comparative} has emerged as a promising tool to solve optimization problems. In addition to performance improvements~\cite{kolotouros_evolving_2022, barron_provable_2024}, the fact that CVaR is computed using bit-strings sampled from a quantum circuit, allows one to work with a customized cost function that can be computed classically and is not necessarily in a quadratic unconstrained binary optimization (QUBO) form. Indeed, we propose a novel conversion approach of the quadratic binary portfolio construction problem that is particularly suited for the quantum sampling-based scheme we use in this study. This eliminates the qubit overhead that would otherwise arise by working with a problem-based diagonal Hamiltonian corresponding to the QUBO problem.

Since VQA algorithms are heuristics and their performance strongly relies on the appropriate choice of ansatz circuits, it is a natural question whether classically hard-to-simulate ansatze can provide better performance than easy-to-simulate ones. Our initial simulation-based results on a 31-qubit quantum circuit indicate that a problem-inspired, harder-to-simulate ansatz, based on the bias-field counterdiabatic optimization scheme~\cite{cadavid_bias-field_2025, romero2024bias} outperforms a more traditional \textit{TwoLocal} ansatz~\cite{qiskit2024}.

Recent works~\cite{sachdeva_quantum_2024, sciorilli2025towards, chandarana2025runtime}  have shown that a classical post-processing of candidate solutions found by the quantum algorithm, based on local search, can often improve the quality of solutions. In this setting, though, it is important to confirm if the contribution of the quantum scheme is actually significant. Our simulation-based study on problems involving $100$+ qubits indicates that the local search alone achieves worse results than in combination with CVaR-VQA. Additionally, we observe that a combined quantum-classical workflow reduces the computational resources when compared with a purely classical local search-based scheme to obtain similar accuracies.

We run 109-qubit hardware experiments on IBM Heron processors. Our results indicate that despite the hardware noise, the raw samples continually improve over an increasing number of iterations and converge towards lower objective values. The raw samples post-processed by the local search scheme further improve the sample quality and provide evidence that a quantum-classical workflow can outperform the purely classical scheme based on local search, as we start tackling relatively large problem sizes.

\section{Problem formulation}

In various industry-relevant scenarios, the problem variables are integer variables $y_i$, where $i=1,...,N$ spans all candidate bonds that may be included in the portfolio. Each $y_i$ represents how many lots of size $\delta_i$ of the $i$th bond are selected for the portfolio. Typically, there exists a minimal and maximal quantity for the bond $i$ if included, so that it is often convenient to define an auxiliary variable $x_i$ stating whether the bond $i$ belongs to the portfolio or not. Bonds are clustered according to different dimensions: for instance, a dimension $D_1 \in \mathcal D$ could be the risk rating, with classes $\{\text{``A+++"}, \text{``A++"}, ...\} = D_1$, and another dimension $D_2$ could be the industry in which the company associated to the bond operates in, say $\{ \text{``Consumer goods"}, \text{``Energy"}, ...\} = D_2$. Moreover, one can associate multiple metrics $ \mu_j$ with $j=1,...,J$ to a bond, such as the ``residual contract duration", the ``key rate duration" or the ``expected excess return". For each class $d^D$ in each dimension $D$, and for each metric $\mu_j$, one can define a target $\tau_{d^{D}, j}$. The optimization problem is then to minimize the average distance of portfolio metrics from the targets:
\begin{equation}
    \argmin_{y} \sum_{D \in \mathcal D} \sum_{d^D \in D} \sum_{j} \left( \tau_{d^{D}, j} - \sum_{i \in d^D} w_{i, D, j} \delta_i y_i \right)^2,
\end{equation}
where $i \in d^D$ spans all bonds belonging to a class $d^D$, and $w_{i, D, j}$ is an appropriate coefficient.
The formulation is typically complemented with some constraints that take the form of linear inequalities. For instance, a typical constraint is the maximal residual cash flow, in the form $$\sum_i p_i \delta_i y_i \leq M,$$
where $p_i$ are normalized prices and $M$ is the budget. Additionally, guardrails are often set to force each of the metric not to deviate excessively from its target:
\begin{multline}
\tau_{d^{D}, j} - \epsilon_{d^{D}, j} \leq \sum_{i \in d^D} w_{i, D, j} y_i \delta_i \leq \tau_{d^{D}, j} + \epsilon_{d^{D}, j}
\\
\forall D \in \mathcal D, \; \forall d^D \in D, \; \forall j.
\end{multline}
The problem allows for a formulation with integer variables, quadratic objectives, and linear constraints. Clearly, it requires additional linear constraints to guarantee the consistency of $y_i$ variables with the corresponding binaries $x_i$.

To make the problem tractable on the current generation of quantum devices with limited physical qubit count and fidelity, we apply some simplifications while still keeping the problem as close as possible to realistic problem conditions. First, we reduce the problem to contain binary variables only. To achieve this, we define $y_i$ to be equal to $c_i x_i$: in other words, if the bond $i$ is included in the portfolio, it is included in the fixed quantity $c_i \delta_i$. Secondly, we reduce the problem size. In order to do so, we select a subset $D'$ of classes in a dimension $D$, and we restrict the set of bonds to those belonging to $D'$. For consistency, we remove all constraints that involve some bonds out of the restricted set. Finally, we manually adjust the cash budget $M$ to conform to the smaller portfolio size.

The original problem poses significant computational challenges only at the scale beyond $1{,}000$ bonds, where a business user may get relatively large optimality gaps from commercial solvers in a typical runtime of $5$ to $10$ minutes. The simplified problem with 109 bonds, on the other hand, can be solved by classical solvers like 
CPLEX or Gurobi to optimality in a few seconds, yielding, in the case of CPLEX, $33$ solutions with one being the proven optimal. However, this is accomplished with some computational effort as indicated by more than $170{,}000$ nodes of the binary search tree examined by the algorithm.

\subsection{Unconstrained formulation}
\label{subsec:unconstrained}
At an abstract level, the simplified problem can be written as
\begin{mini}
{x \in \{0,1\}^n}
{x^T Q x}
{}
{}
\addConstraint{A x}{\leq b}{}
\end{mini}
with some $Q \in \mathbb R^{n \times n}$, $A \in \mathbb R^{m \times n}$, and $b \in \mathbb R^m$.

It should be noted that during the conversion of the problem to the binary unconstrained form, the inequality constraints are usually handled by the introduction of slack variables, which in turn leads to an increase in the number of binary variables. Few recent studies~\cite{montanez2024unbalanced, takabayashi2025subgradient} however propose to handle such constraints without the introduction of slack variables. In this work, we propose an equivalent unconstrained form of the problem as
\begin{mini}
{x \in \{0,1\}^n}
{x^T Q x + 1^T \max\{0, s \odot (Ax-b)\},}
{}
{}
\end{mini}
where $s \in \mathbb R^m$ is an appropriate rescaling factor for constraints and $\odot$ is the element-wise product. The values of vector $s$ must be set large enough so that any constraint violation must always increase the objective function. As the term $1^T \max\{0, s \odot (Ax-b)\}$ is always non-negative, if values of the $s$ vector are large enough, the term will be zero for an optimal solution.

\section{Methods}

\subsection{Sampling-based CVaR-VQA}
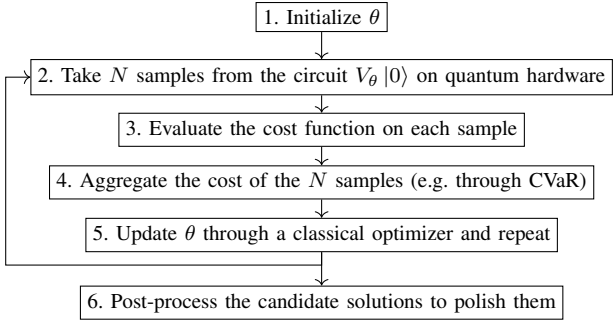
\begin{figure}
    \centering \footnotesize
\begin{tikzpicture}
\node[draw] (init) at (0,0)  {1. Initialize $\theta$};
\node[draw] (sample) at (0,-.8)  {2. Take $N$ samples from the circuit $V_{\theta} \ket{0}$ on quantum hardware};
\node[draw] (cost) at (0,-1.5) {3. Evaluate the cost function on each sample};
\node[draw] (aggregate) at (0,-2.2) {4. Aggregate the cost of the $N$ samples (e.g. through CVaR)};
\node[draw] (update) at (0,-2.9) {5. Update $\theta$ through a classical optimizer and repeat};
\node[draw] (polish) at (0,-3.8) {6. Post-process the candidate solutions to polish them};

\draw [->] (init) -- (sample);
\draw [->] (sample) -- (cost);
\draw [->] (cost) -- (aggregate);
\draw [->] (aggregate) -- (update);
\draw [->] (update) -- (polish);
\draw [->] (update) |- (-4.2, -3.3) |- (sample);;
\end{tikzpicture}
    \caption{The workflow of sampling-based VQA. Only step~2. involves a quantum processing unit (QPU).}
    \label{fig:workflow}
\end{figure}

Suppose we are given a binary unconstrained optimization problem with $n$ variables, and with a cost function $f: \{0,1\}^n \to \mathbb R$.
In Fig.~\ref{fig:workflow}, we provide the high-level workflow of the sampling-based CVaR-VQA scheme. Similar to a variational quantum scheme, we use a parametrized circuit $V(\theta)$ characterized by parameter vector $\theta$, where $\theta \in [0,2 \pi)^m$. At the beginning, the  circuit parameter is initialized to a value $\theta_0$. Then the variational circuit is executed, followed by measurement in the computational basis to collect $N$ sample bit-strings $\{x_i\}_{i=1}^N$, where $x_i \in \{0,1\}^n$. The cost function $f(x_i)$ is then computed over the samples $x_i$. The computed cost function values $\{f(x_i)\}_{i=1}^N$ are then aggregated using CVaR(${\alpha}$) to compute an overall score $s(\theta)$. The aggregation based on CVaR(${\alpha}$) under these settings can be seen as the average of lower $\alpha$-tail of the distribution of the cost function values. When $\alpha=1$, it amounts to computing the average cost function values over all the samples, whereas when $\alpha \to 0$, then only the lowest cost sample is considered. When setting $\alpha$ in the range $(0, 1]$, then CVaR is known to lead to better convergence in quantum optimization~\cite{barkoutsos_improving_2020, kolotouros_evolving_2022} and shows to be more robust against hardware noise~\cite{barron_provable_2024}. Subsequently, a classical optimizer is used to update the circuit parameters at every iteration in order to find the optimal parameter $\theta_{\mathrm{opt}}$ that minimizes $s(\theta)$. The samples collected from the optimized circuit are further post-processed using a local-search scheme.

Compared to other quantum optimization approaches, the sampling-based VQA offers some important benefits, since the calculation of the cost is performed classically and its applicability is not restricted to QUBOs, but rather allows for arbitrarily complex objective functions such as higher-order and non-polynomial objectives. As described in Subsec.~\ref{subsec:unconstrained}, it allows us to introduce problem constraints without any qubit overhead. As a consequence, we use only one qubit per bond. Finally, sampling-based VQA collects samples along iterations, which then can be post-processed classically.

\subsection{Ansatz}
\begin{figure}
    \centering \footnotesize
\begin{tikzpicture}
\node[above] at (0,0) {
    \begin{quantikz}[column sep=1mm, row sep=2mm]
    \lstick{\ket{0}} & \gate{} &[2mm] \gate[3]{} \gategroup[3,steps=2,style={dashed, inner
    sep=1pt, color=blue}, label style=blue]{$r$ repetitions} & \gate{} &[2mm] \meter{} \\
    \setwiretype{n} & \vdots & & \vdots & \vdots \\
    \lstick{\ket{0}} & \gate{} &  & \gate{} & \meter{}
    \end{quantikz}
};
\node[above] at (3,0) {
    \begin{quantikz}[column sep=1mm, row sep=1mm]
    & \gate[2]{} & & \\
    & & \gate[2]{} & \\
    & \gate[2]{} & & \\
    & & \gate[2]{} & \\
    & \gate[2]{} & & \\
    & & & 
    \end{quantikz}
};
\node[above] at (5.4,0) {
    \includegraphics[width=2.2cm, trim={8cm 11cm 6cm 2cm}, clip]{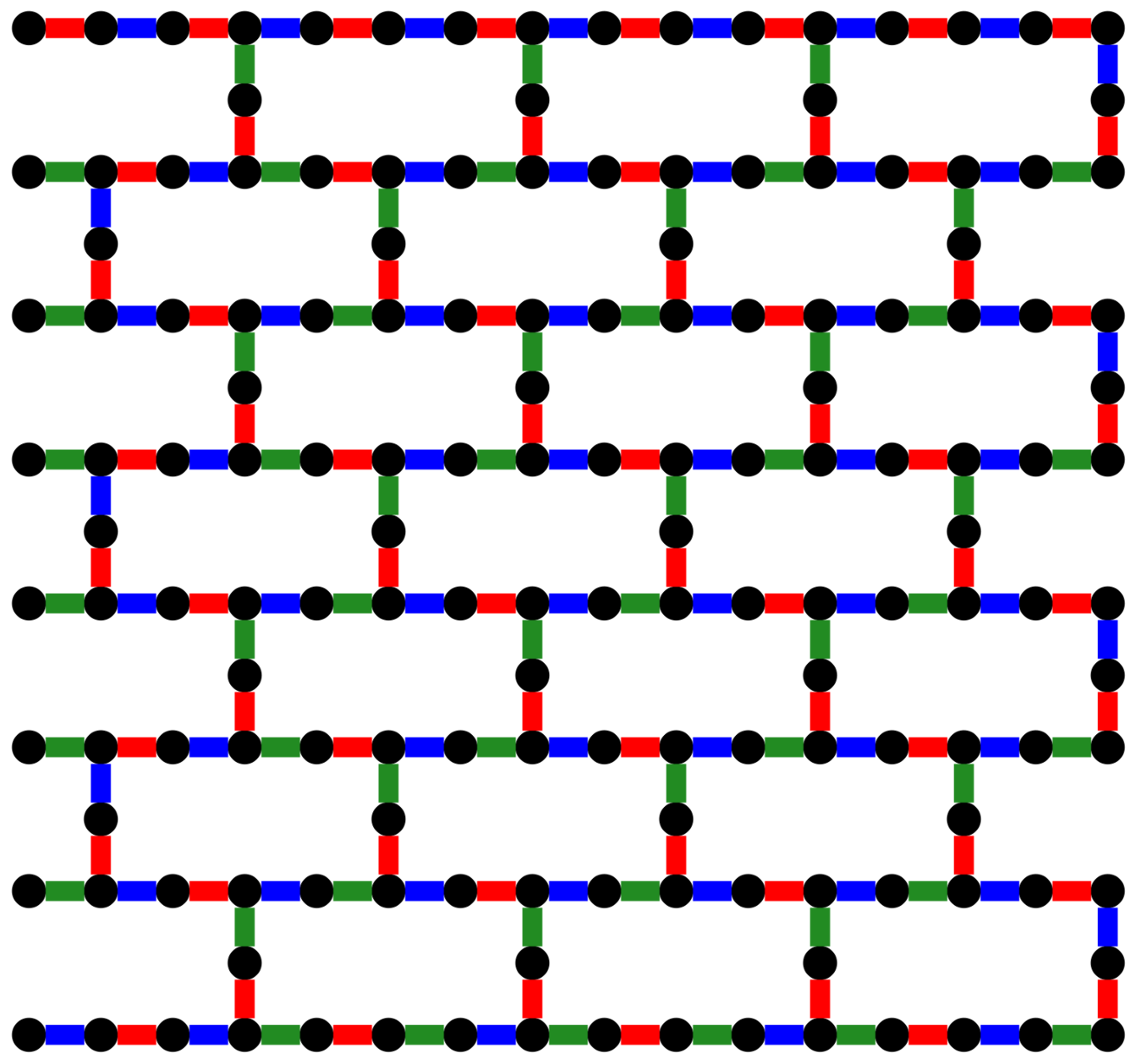}
};
\node at (0,-.3) {
    \textbf{a)}
};
\node at (3,-.3) {
    \textbf{b)}
};
\node at (5.4,-.3) {
    \textbf{c)}
};
\end{tikzpicture}

    \caption{a)~General design of the ansatz $V(\theta)$. b)~Bilinear entanglement. c)~The colored entanglement is defined over a 3-label coloring on the IBM's heavy hex chip topology.}
    \label{fig:ansatz}
\end{figure}
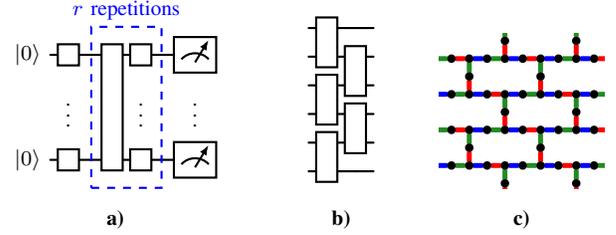

\begin{table*}[t]
    \centering
    \renewcommand{\arraystretch}{1.1}
    \begin{tabular}{cccccccc}
\hline
Entanglement structure	&	Ansatz	&	Depth	&	Gate count	&	Two-qubit gate count	&	Param count	&	Simulator runtime (1k shots) \\
\hline
bilinear	&	TwoLocal	&	17	&	1525	&	216	&	327	&	5 s	\\
color	&	TwoLocal	&	19	&	1555	&	246	&	327	&	3 min	\\
bilinear	&	BFCD	&	57	&	4220	&	648	&	434 &	 16 min	\\
\hline
\\[.2mm]
    \end{tabular}
    \caption{Circuit characteristics of the 109-qubit circuits transpiled on \texttt{ibm\_marrakesh}}
    \label{tab:sizes}
\end{table*}

As shown in Fig.~\ref{fig:ansatz}a, our ansatz $V(\theta)$ consists of multiple repetitions $j=1,...,r$ of alternating single-qubit rotation gates and two-qubit entanglement layers. We use two different two-qubit gate connectivity for the entanglement layer, which we refer as bilinear and colored. For the bilinear entanglement map, see Fig.~\ref{fig:ansatz}b, the two-qubit gates first entangle qubits with an odd number index followed by qubits with an even number index. We set the repetitions $r=2$. The transpiled depth  in this case remains constant, given the linear chain of $n$ qubits that completely fits the hardware graph. The ``colored'' entanglement structure, instead, is designed to mirror the device topology. We first apply a 3-label coloring to the native heavy-hex connectivity on the IBM devices, see Fig.~\ref{fig:ansatz}c, and then we entangle equally colored qubits in parallel, by means of 2-qubit gates. Consequently, the depth of the entanglement layer is~3 before transpilation, and it remains independent of $n$ after transpilation. If the number $n$ of variables is lower than the number of qubits on the device, we iteratively remove one low-degree node from the topology graph until the desired size is reached.

Along with the two entanglement maps, we also consider here two different ansatze characterized by different gate sets. One of the ansatz we employ is the \textit{TwoLocal} ansatz characterized by parametrized $R_Y$ as single-qubit rotations and controlled-$Z$ as two-qubit gates. The second ansatz, we refer as BFCD, is inspired by bias-field counterdiabatic optimization~\cite{cadavid_bias-field_2025, romero2024bias}. The original BFCD circuit, in the impulse regime, is approximated by alternating $R_Y$ rotations and two-qubit rotation gates $R_{YZ} R_{ZY}$ (see Fig.~S1 in the Supplementary Information of Ref.~\cite{cadavid_bias-field_2025}). Therein, the rotation angles are analytically computed, and the two-qubit gates are applied to all qubit pairs. Instead, our proposed BFCD ansatz parametrizes the rotation angle in each gate. To decrease the number of parameters and improve trainability, we use the same parameter for $R_{YZ}$ and the corresponding $R_{ZY}$, inspired by the fact that in the original work, those angles have the same value when analytically computed.

Table~\ref{tab:sizes} summarizes the key metrics for the two entanglement maps and the two ansatze for the 109-qubit circuit transpiled on the IBM Heron processor \texttt{ibm\_marrakesh}. The number of parametrized gates in the BFCD ansatz is 32\% more than the \textit{TwoLocal} ansatz, and hence it is expected to require more iterations to get to same converged cost values compared to the \textit{TwoLocal} ansatz. The circuits exhibit up to 648 two-qubit gates and around 4000 total gates. Additionally, we compare simulation runtimes measured using the Matrix Product State (MPS) simulator~\cite{qiskit2024} on an Apple M1 Pro chip with 16GB RAM. We observe the circuit with the color entanglement map to take more time to simulate. Notably, the BFCD ansatz with bilinear entanglement map takes the longest to simulate. This increased time in simulating these circuits using the MPS simulator can be attributed to the increase in the degree of entanglement present in the circuits.

\subsection{Classical optimizer: NFT}
As a classical optimizer, we choose the Nakanishi-Fujii-Todo (NFT) method~\cite{nakanishi_sequential_2020}, a gradient-free technique that exhibits relatively better convergence properties in the presence of hardware noise~\cite{oliv_evaluating_2022, singh_benchmarking_2023}. The algorithm proceeds by taking a step for each variational parameter, one at a time in a randomized order. At each step, the partial derivative along the direction of the chosen parameter is estimated with a 3-point numerical approximation, thus requiring 3 circuit evaluations. One epoch is complete when all parameters have been updated once. Then, the parameter order is reshuffled, and a new epoch is run, iteratively.
Additionally, during hardware runs, we employ a parameter cut-off~\cite{romero2024bias, cadavid_bias-field_2025, kumar2025towards}, where gates with parameter below the threshold $0.06$ are ignored. Moreover, if the set of parameters, for an iteration, is not updated by the NFT optimizer, we do not rerun the evaluation of $f(\theta)$.

\subsection{Post-processing with local search}
Previous work~\cite{sachdeva_quantum_2024, chandarana2025runtime, kumar2025towards} showed that local search-based post-processing can substantially improve the performance of the optimization scheme. In this work, we examine, in detail, the effect of local search on the samples obtained from the quantum scheme. Following Ref.~\cite{sachdeva_quantum_2024}, the local search starts from one candidate solution. For each candidate solution, we flip one variable at a time, in a randomized order, until we either improve the cost, or until we have considered all the variables. When a better cost is found, the local search restarts from this new point. %
For the purpose of this analysis, we post-process all the sampled bit-strings at every iteration, but our results confirm that it is possible to restrict only to the last iterations. It should be noted that the CVaR values reported here are computed using samples from the quantum scheme with no post-processing.

\section{Experimental results}
We run quantum circuits using the MPS simulator (with no truncation of the bond dimension) as well as conduct hardware experiments on the IBM Quantum Heron processors. The circuit are transpiled using Qiskit Primitive V2 transpiler with the optimization level set to 3. Dynamical decoupling is employed for error suppression using the default $XX$ sequence. The resulting transpiled circuit is then executed on the quantum processor using the sampling primitive of Qiskit, with $N_{shots}=2^{13}$ measurements at each iteration. In all of our runs, each circuit parameter is initialized to $\pi/3$. %

\subsection{31-qubit MPS simulation}
\begin{figure}[t]
    \centering \footnotesize
    \begin{picture}(\textwidth, 130)
        \put(10,0){\includegraphics[width=.95\linewidth]{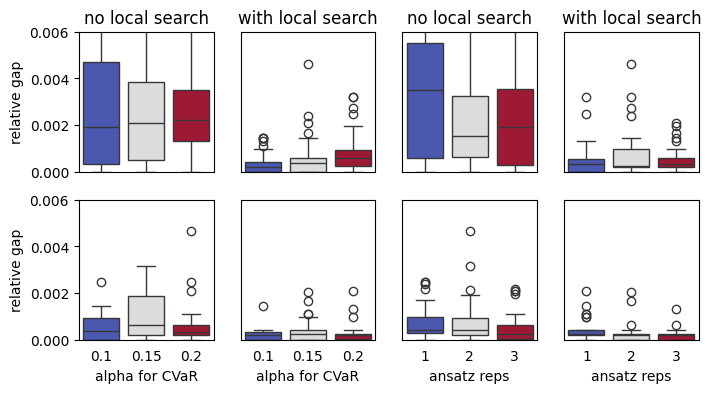}}
        \put(0,75){a)}
        \put(0,18){b)}
    \end{picture}
    \caption{Tuning of $\alpha$ and $r$ for a) the \textit{TwoLocal} bilinear ansatz and b) the BFCD bilinear ansatz, based on 31-qubit experiments on simulator.}
    \label{fig:tuning}
\end{figure}

We first use MPS-based simulation on a 31-qubit problem to assess the performance of the scheme for increasing values  of CVaR $\alpha$ and the circuit repetitions $r$. Fig.~\ref{fig:tuning} contains results, for the \textit{TwoLocal} ansatz, over 90 experiments (10 independent repetitions of each experiment with $\alpha \in \{ 0.1, 0.15, 0.2\}$ and $r \in \{ 1,2,3 \}$). We measure the solution quality in terms of relative gap $\gamma = \frac{\left| F(\theta_f)_{\text{low}} - F_0 \right|}{|F_0|}$ to the optimal solution (optimal solution was found by CPLEX). Fig.~\ref{fig:tuning} indicates relative gaps obtained with and without local search. In the case of local search, we only process samples from the last 20 iterations. %
 Fig.~\ref{fig:tuning} indicates that $\alpha = 0.1$, clearly yields a better relative gap overall. We also observe that BFCD ansatz compared to the \textit{TwoLocal} ansatz provides better results with increasing repetitions. As BFCD ansatz is harder to simulate classically using the MPS simulator (see Table~\ref{tab:sizes}), our result provides evidence that one can find a ``useful'' class of quantum circuits with better convergence properties that may have large classical simulation overhead. 
 
 Based on the results, we set $\alpha =0.1$ for the subsequent hardware runs. Additionally, we set repetitions  $r=2$, not only because it offers good results in terms of the relative gap, but also because it provides a limited, yet non-trivial, gate depths (see Table~\ref{tab:sizes}), at which we can expect good results also on the hardware.

\subsection{109-qubit experiment}
\begin{figure}[t]
    \centering \footnotesize
    \includegraphics[width=\linewidth]{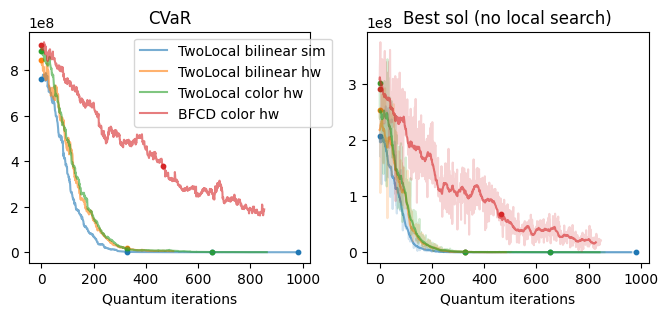}
    \caption{Comparison of ansatz convergence on hardware and simulator, without local search for the 109-qubit problem. Left: CVaR. Right: best cost found at each iteration (shaded area) and respective moving average (line). Dots mark the NFT epochs.}
    \label{fig:109-no-ls}
\end{figure}

    \label{fig:109-ls}

\begin{figure}[t]
    \centering \footnotesize
    \includegraphics[width=\linewidth]{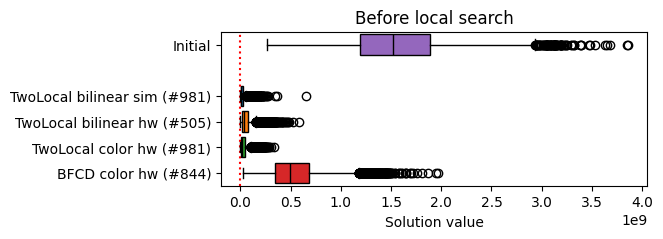}
    \caption{Distributions of the objective values sampled from each ansatz at the end of the optimization, benchmarked against the `Initial' distribution (i.e. iteration 1 on simulator of the the \textit{TwoLocal} bilinear). Numbers in brackets indicate the iterations performed in the respective run.}
    \label{fig:109-sampledist-before}
\end{figure}

\begin{figure}[t]
    \centering \footnotesize
    \includegraphics[width=\linewidth]{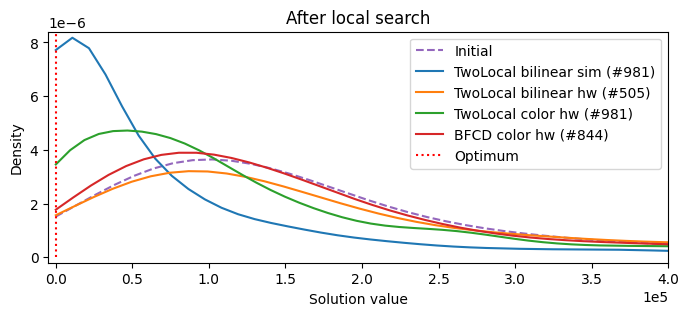}
    \caption{Kernel Density Estimate (KDE) plot for the objective value sampled from each ansatz at the end of the optimization and then processed with local search, benchmarked against the `Initial' distribution (i.e. iteration 1 on simulator of the \textit{TwoLocal} bilinear). Numbers in brackets indicate the iterations performed in the respective run.}
    \label{fig:109-sampledist-after}
\end{figure}

We run 109 qubit experiments involving each of the following three ansatze: the \textit{TwoLocal} bilinear and color on the IBM Heron processors \texttt{ibm\_marrakesh} and \texttt{ibm\_fez}, respectively, and the BFCD on \texttt{ibm\_marrakesh}. As the computational cost of MPS-based simulator grows exponentially with entanglement~\cite{vidal2003efficient}, we limit our simulations to the \textit{TwoLocal} bilinear ansatz (see Table \ref{tab:sizes}).

Fig.~\ref{fig:109-no-ls} shows the evolution of the sample-based CVaR and best objective value, with iteration, computed over raw samples. The best solution found among the raw samples during the final iteration has close proximity (small relative gap) to the optimal solution (found by CPLEX). The simulator requires fewer iterations to find good solutions, implying that hardware noise slows the process of tuning variational parameters. However, we observe that even in presence of hardware noise, sample quality keeps improving with more iterations. The plot indicates that with enough number of iterations, hardware samples will reach sufficient convergence. As the total runtime of the scheme depends on the number of iterations $N_{\mathrm{iter}}$ and circuit execution time $T_c$, a larger number of $N_{\mathrm{iter}}$ required on the hardware may still recover speed-ups against classical simulation schemes if the time $T_c$ is small on the real hardware. For instance, if hardware requires 2$\times$ the number of iterations but is then  10$\times$ faster than simulators time-wise, then still quantum hardware will be 5$\times$ faster in the overall time taken to run the scheme.

In Fig.~\ref{fig:109-sampledist-before}, we show the distribution of the cost function evaluated on the sampled candidate solutions, without local search. The distribution from the circuit at first iteration, characterized by initial parameters, is observed to show a peak around $1.5\cdot10^9$, significantly far from the optimal solution. The plot indicates that after the circuits are variationally trained, they exhibit distributions that approach the actual optimum (dashed red line), for all ansatze. Note that the mean sampled cost function values using the BFCD ansatz are relatively far from optimal compared to the other circuits which can be explained by the slower convergence observed in Fig.~\ref{fig:109-no-ls} for the BFCD ansatz.
In Fig.~\ref{fig:109-sampledist-after}, we plot the cost function value distribution computed over the hardware samples, post-processed by the local search. We observe that the sampling probabilities, particularly for the simulator and the \textit{TwoLocal} color hardware run, significantly improve due to local processing. However, the post-processed distribution from the BFCD circuit exhibits relatively lower sampling probabilities compared to other ansatz and is similar to the post-processed initial distribution. This indicates that sufficiently converged raw samples from the quantum scheme are essential to gain improvements from the local-search. A local-search-based post-processing may fail if the raw samples are relatively far from the optimal solution. Hence, these results provide evidence that a quantum-classical workflow can significantly yield better performance that cannot be achieved by a purely classical local-search-based scheme.

The best performance after local search is achieved by the \textit{TwoLocal} bilinear ansatz ($0.49\%$ optimization gap), followed by the \textit{TwoLocal} color ($0.55\%$) and BFCD color ($0.91\%$). 
In the simulator run, the gap is $0.39\%$.

\section{Conclusions}
In this work, we examined the feasibility of using CVaR-VQA combined with a local search scheme to solve a simplified but realistic portfolio construction problem for a specific bond ETF instance. Our 31-qubit MPS simulation runs indicate that circuits with a higher degree of entanglement may yield better solutions. As such circuits would take more time to classically simulate, they hint towards the possibility of a regime where quantum schemes can truly excel over purely classical and simulation-inspired schemes. However, more work needs to be done to further establish these trends.

We observe that the 109 qubit noiseless simulations and noisy hardware runs achieve encouraging results. The hardware run, despite noise, is able to evolve continuously towards better solutions as the circuit is trained. Post-processing of the samples using local search leads to better quality samples. Our results provide evidence that a combined quantum-classical workflow can bring samples close to the optimal, which is something that cannot be achieved by only using the local search method.

In order for quantum methods to provide an advantage over classical, though, it is essential to scale the problem size to a much larger regime, where classical solvers struggle. As the portfolio construction problem studied in this work becomes classically hard only at large scale, it is important to 
ensure that the quantum techniques can scale up at an affordable cost. A detailed analysis of the scaling of the proposed technique is beyond the scope of the current work and likely out of reach with modern empirical settings, due to the limited maturity of quantum hardware. Indeed, we believe that little can be inferred about the runtime at large scale, from experiments of small to mid problem size, given the heuristic nature of the methods.
Recent studies~\cite{fuller2024approximate, sciorilli2025towards} provide a path towards tackling larger-scale problems. %
As a potential limitation inherent to the CVaR-VQA scheme adopted here, the quantum-classical training of quantum circuits over multiple iterations, leads to several circuit executions on the hardware, which grow with the number of parameters and hence with the problem size. In order to avoid such a situation, one can adopt a classical-only training mode where parameter transfer techniques~\cite{shaydulin2023parameter} can be used or training is only carried out on the classical compute nodes and the optimized circuit is then run on the hardware to collect samples~\cite{kumar2025towards}. Our future work will investigate these techniques in more detail.

\section*{Acknowledgment}
Authors thank Paul M Malloy and Brian Jaffe for their support throughout the project, as well as Stefan Woerner for helpful technical discussions.

\bibliographystyle{IEEEtran}
\bibliography{biblio}

\end{document}